**Twistsonics: engineering acoustic topological textures in moiré sound lattices**

*Wen-Yu Wang, Shuai Liu, Xiang-Yuan Xu, Qing Tong, Hao Ge,* Yijie Shen,* and Ming-Hui Lu**

Wen-Yu Wang, Hao Ge, Ming-Hui Lu
National Laboratory of Solid State Microstructures & School of Advanced Manufacturing Engineering, Nanjing University, Suzhou 215163, China

Wen-Yu Wang
School of Physics, Nanjing University, Nanjing, Jiangsu 210093, China.

Shuai Liu, Xiang-Yuan Xu, Hao Ge, Ming-Hui Lu
College of Engineering and Applied Sciences, Nanjing University, Nanjing, Jiangsu 210093, China.

Qing Tong, Yijie Shen
Centre for Disruptive Photonic Technologies, School of Physical and Mathematical Sciences & School of Electrical and Electronic Engineering, Nanyang Technological University, Singapore, Singapore.

Ming-Hui Lu
Jiangsu Key Laboratory of Artificial Functional Materials & Jiangsu Physical Science Research Center, Nanjing, Jiangsu 210093, China.

**E-mail:** Hao Ge (haoge@nju.edu.cn), Yijie Shen (yijie.shen@ntu.edu.sg), Ming-Hui Lu (luminghui@nju.edu.cn)

**Funding:** The National Key R&D Program of China (Grants No. 2023YFA1406904 and Grants No. 2021YFB3801801), the National Natural Science Foundation of China (Grants No. 52250363 and Grants No. 52203358), the Natural Science Foundation of Jiangsu Province (Grants No. BK20232048 and Grants No. BK20233001), Singapore Ministry of Education (MOE) AcRF Tier 1 grants (Grant Nos. RG157/23 and RT11/23), Singapore

Agency for Science, Technology and Research (A*STAR) MTC Individual Research Grants (Grant No. M24N7c0080), and Nanyang Assistant Professorship Start Up grant.

**Keywords:** acoustic moiré superlattice, skyrmion bags, topological sound textures, spoof surface acoustic waves, acoustic metasurface

**Abstract**

Moiré superlattices formed by twisting periodic systems provide a powerful platform for emergent topological phenomena, but their use for programming real-space topology in acoustic wave fields remains largely unexplored. Here we report a phase-controlled spoof surface acoustic wave platform for constructing moiré topological textures in the acoustic particle-velocity field. On a perforated acoustic metasurface, two twisted skyrmion lattices are synthesized and superposed, yielding acoustic skyrmion bags with controllable topological numbers and geometries. The twist angle and rotation center control the scale and configuration of the bags, enabling deterministic reshaping of the composite texture. We further introduce controlled defects to assess the defect tolerance of the moiré skyrmion bags. The skyrmion bags retain their composite topology over a finite range of defect densities, and comparison with an untwisted single-layer skyrmion lattice suggests enhanced stability of the enclosed skyrmion cluster relative to the single-layer reference at higher defect densities. The same programmable platform also supports transitions from skyrmion lattices to meron lattices and enables twist-induced meron clusters. These findings establish acoustic moiré superlattices as a reconfigurable platform for engineering robust real-space topological textures, with potential applications in topology-guided acoustic manipulation and information encoding.

(Author 1 and Author 2 contributed equally to this work.)

## 1. Introduction

Moiré superlattices arise when two periodic lattices are twisted or slightly mismatched,[1-5] giving rise to a long-wavelength modulation that reorganizes the effective periodicity and interaction landscape.[2,6] In electronic materials, such moiré potentials have enabled band-structure engineering,[1-3,5] including the emergence of flat bands[2] and correlated electronic phases in twisted bilayer systems,[6-8] forming the foundation of twistronics. Related moiré concepts have since been extended to photonic,[9-12] plasmonic,[13-15] hydrodynamic[16] and acoustic systems,[17,18] where they provide new means to tailor dispersion, localization and real-space field distributions.[9,13,18,19] These studies show that twist-induced interference can organize wave fields beyond the scale of the constituent lattices. However, using acoustic moiré superlattices to program real-space topological textures remains largely unexplored.

Topological textures such as skyrmions and merons[20-24] are vector-field configurations with nontrivial orientational winding and quantized topological charge.[21,22,25,26] Their robustness against perturbations[21,27] and capacity to encode structured field information[22,28] have stimulated broad interest across condensed-matter[29-33] and classical-wave systems.[24,26,34-43] In acoustics, skyrmionic and related topological textures have only recently begun to emerge.[44-49] Acoustic wave fields offer distinct advantages for such studies because structured fields can be programmably synthesized, while the underlying vector particle-velocity field can be directly reconstructed for quantitative characterization of real-space topology. However, existing demonstrations are typically tied to fixed structural designs, limiting continuous and reversible reconfiguration of real-space topological textures.

Here we introduce a reconfigurable acoustic platform that uses moiré interference to program real-space topological textures. Based on phase-controlled spoof surface acoustic waves, we synthesize two-dimensional skyrmion lattices and superpose them with a prescribed relative rotation, such that the twist angle and rotation center serve as independent geometric control parameters. This twist-driven bilayer configuration generates acoustic moiré textures that support composite topological states, with acoustic skyrmion bags exhibiting controllable topological numbers and geometries. By introducing controlled defects, we show that the skyrmion bags retain their composite topology over a finite range of defect densities. Comparison with an untwisted single-layer skyrmion lattice further suggests enhanced stability of the enclosed skyrmion cluster relative to the single-layer reference. The same platform also supports phase-driven transitions from skyrmion lattices to meron lattices and the formation of twist-induced meron clusters. These results establish moiré interference as a practical route for engineering real-space topological textures in acoustic wave fields.

## 2. Results and Discussion

### 2.1. Twist-engineered acoustic moiré platform

Superposing two periodic lattices with a relative twist generates a long-wavelength moiré pattern in real space (see **Figure 1a**). When the constituent lattices are vector acoustic skyrmion lattices, the resulting moiré interference gives rise to composite topological textures, including skyrmion bags, in which a skyrmion cluster is enclosed by an oppositely oriented boundary.[13,15] To implement this concept experimentally, we use a phase-programmed spoof surface acoustic wave (SSAW) platform based on a perforated acoustic metasurface with hexagonally arranged subwavelength apertures (see Figure 1b). In our platform, the lattices are encoded in the particle-velocity field of SSAWs. The metasurface supports SSAWs that are strongly confined near the surface and carry a spatially varying vector particle-velocity field with local rotational character. Dropping the time-dependence factor, the velocity field of SSAWs takes the form:

$$\vec{v} = \frac{p_0}{\rho_0 \omega} \begin{pmatrix} k_x \\ k_y \\ i\tau \end{pmatrix} e^{i(k_x x + k_y y) - \tau z} \tag{1}$$

where $p_0$ is the wave amplitude, $\rho_0$ is the ambient air density, and $\omega$ is the angular frequency. $k_x$ and $k_y$ denote the in-plane wavevector components, while the out-of-plane component is $k_z = i\tau$. The SSAW dispersion relation satisfies $k_x^2 + k_y^2 - \tau^2 = (\omega/c_0)^2$, where $c_0$ is the speed of sound in air under ambient conditions.

The experimental configuration is shown schematically in Figure 1c. Acoustic topological textures are generated by the surrounding six speaker arrays, and the resulting particle-velocity field is measured above the metasurface. The surrounding speaker arrays excite three pairs of counter-propagating SSAWs along the $\theta = 0°$, $120°$, and $240°$ directions, thereby forming three standing-wave components. When these components have identical spatial phases, their interference produces a hexagonal acoustic skyrmion lattice (see Figure 1d), described by:

$$\vec{v} = \frac{2p_0}{\rho_0 \omega} e^{-\tau z} \sum_{n=0}^{2} \begin{bmatrix} k_{n,x} \sin(\vec{k}_n \cdot \vec{r} + \varphi_n) \\ k_{n,y} \sin(\vec{k}_n \cdot \vec{r} + \varphi_n) \\ \tau \cos(\vec{k}_n \cdot \vec{r} + \varphi_n) \end{bmatrix}, \tag{2}$$

where $\varphi_n$ are the spatial phases of the standing-wave components, $\vec{k}_n = \sqrt{k_x^2 + k_y^2}\,[\cos(2n\pi/3), \sin(2n\pi/3)]$, and $\vec{r} = (x, y)$.

A twist degree of freedom is introduced through direction modulation of the excited SSAWs. By tailoring the driving phases of the speaker array on each edge, we generate two SSAWs tilted by $-\alpha/2$ and $+\alpha/2$ with respect to the normal of the corresponding metasurface edge, thereby controlling the orientation of the generated skyrmion lattice. Two skyrmion lattices rotated by $-\alpha/2$ and $+\alpha/2$ can thus be synthesized simultaneously on the same metasurface (see Figure 1e). Their prescribed relative rotation defines the twist angle $\alpha$, and their superposition yields a twisted-bilayer interference field within a single acoustic plane. This moiré interference reshapes the underlying vector texture and supports skyrmion bags, as highlighted in the enlarged view in Figure 1f. The corresponding velocity field is described by:

$$\vec{v} = \frac{2p_0}{\rho_0 \omega} e^{-\tau z} \sum_{n=0}^{2} \begin{bmatrix} k_{n,1,x} \sin(\vec{k}_{n,1} \cdot \vec{r} + \varphi_{n,1}) + k_{n,2,x} \sin(\vec{k}_{n,2} \cdot \vec{r} + \varphi_{n,2}) \\ k_{n,1,y} \sin(\vec{k}_{n,1} \cdot \vec{r} + \varphi_{n,1}) + k_{n,2,y} \sin(\vec{k}_{n,2} \cdot \vec{r} + \varphi_{n,2}) \\ \tau \cos(\vec{k}_{n,1} \cdot \vec{r} + \varphi_{n,1}) + \tau \cos(\vec{k}_{n,2} \cdot \vec{r} + \varphi_{n,2}) \end{bmatrix} \quad (3)$$

where $\varphi_{n,1}$ and $\varphi_{n,2}$ are the spatial phases of the standing-wave components forming the two skyrmion lattices, respectively. $\vec{k}_{n,1} = \sqrt{k_x^2 + k_y^2}\,[\cos(2n\pi/3 - \alpha/2), \sin(2n\pi/3 - \alpha/2)]$, and $\vec{k}_{n,2} = \sqrt{k_x^2 + k_y^2}\,[\cos(2n\pi/3 + \alpha/2), \sin(2n\pi/3 + \alpha/2)]$.

In this construction, the twist-induced texture is governed by two geometric parameters: the twist angle $\alpha$ and the rotation center. The twist angle sets the characteristic scale of the moiré superlattice, whereas the rotation center determines the relative alignment of the two lattices across the plane, so that varying its position produces different skyrmion-bag configurations. In the experiment, the rotation center is referenced to the geometrical center of the metasurface. By adjusting the spatial phases $\varphi_{n,1}$ and $\varphi_{n,2}$, we translate the two constituent skyrmion lattices relative to the metasurface center, which effectively shifts the rotation center relative to the lattice pattern. Together, $\alpha$ and the rotation center provide independent geometric controls for reconfiguring real-space acoustic topological textures.

To identify the resulting textures quantitatively, we normalize the particle-velocity field $\vec{v}$ as $\vec{n} = \vec{v}/|\vec{v}|$ and compute the skyrmion number:

$$S = \frac{1}{4\pi} \int \vec{n} \cdot \left( \frac{\partial \vec{n}}{\partial x} \times \frac{\partial \vec{n}}{\partial y} \right) dxdy. \quad (4)$$

The integrand $s = \frac{1}{4\pi} \vec{n} \cdot (\frac{\partial \vec{n}}{\partial x} \times \frac{\partial \vec{n}}{\partial y})$ defines the skyrmion number density. Textures carrying integer topological charge are identified as skyrmions, whereas those with half-integer charge correspond to merons. For a skyrmion bag, the enclosed skyrmion cluster contributes an integer charge $S_{\text{cluster}} = N$, while the oppositely oriented outer boundary contributes an additional $-1$. The total charge of the composite object is therefore $S_{\text{bag}} = S_{\text{cluster}} - 1$.

Experimentally, the three-dimensional particle-velocity field above the metasurface is mapped using a particle-velocity sensor mounted on a translation stage, with a scan step of 5 mm along both the $x$ and $y$ directions. The data-acquisition module records the amplitude and phase of the acoustic particle velocity, while a synchronized reference signal enables retrieval of the complex $v_x$, $v_y$, and $v_z$ components at each spatial point. These measurements are then used to reconstruct the full vector velocity field and extract the corresponding topological textures. Owing to the harmonic time dependence of the field, the textures are visualized at the instant when the out-of-plane component $v_z$ at the skyrmion centers reaches its positive maximum.

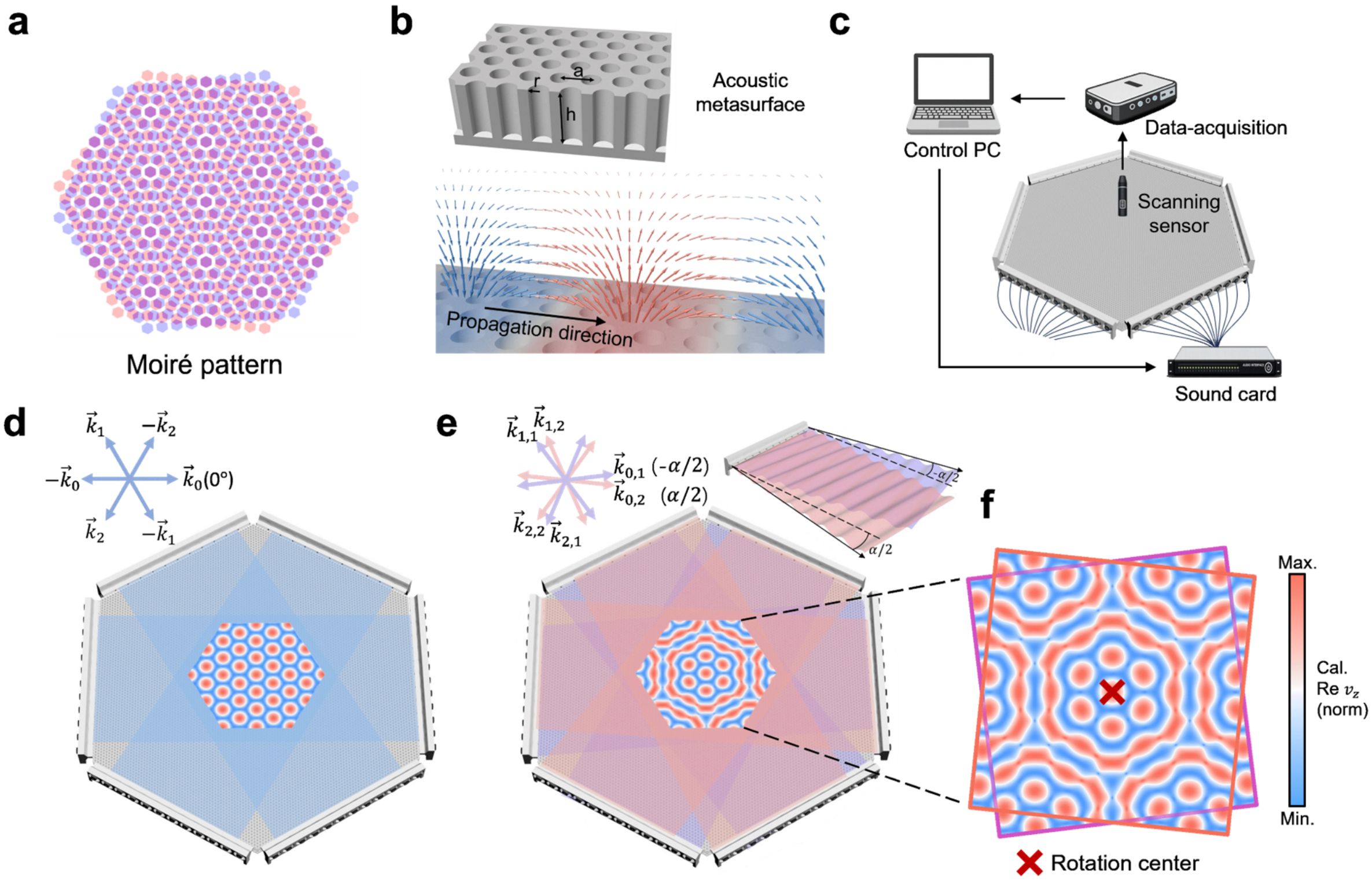

**Figure 1. Twist-engineered acoustic topological textures realized using phase-programmed spoof surface acoustic waves.** (a) Schematic of a moiré pattern formed by superposing two hexagonal lattices with a relative twist. (b) Acoustic metasurface with hexagonally arranged subwavelength apertures for supporting spoof surface acoustic waves (SSAWs), with a lattice constant $a = 10$mm, aperture radius $r = 3.75mm$, and aperture height $h = 30mm$. The lower panel shows the propagation of an SSAW and the corresponding particle-velocity field distribution. (c) Schematic of the experimental measurement setup. Acoustic topological textures are generated by an acoustic phased array driven by multichannel sound cards, and the acoustic particle-velocity field is mapped on a scanning plane above the metasurface using a velocity sensor. (d) Formation of a skyrmion lattice on the metasurface by three pairs of counter-propagating SSAWs. The inset shows the corresponding six in-plane SSAW wave vectors. (e) Generation of twisted bilayer skyrmion lattices by direction modulation. Each speaker array on a metasurface edge produces two SSAWs tilted by $-\alpha/2$

and $\alpha/2$ with respect to the normal of the corresponding edge, thereby generating two skyrmion lattices rotated by $-\alpha/2$ and $\alpha/2$. Their superposition yields an interference field supporting skyrmion bags. The insets show the direction-modulation scheme and the corresponding wave-vector configuration. (f) Enlarged view of the resulting moiré topological texture. The colored square frames indicate the two constituent skyrmion lattices, and the red cross marks the rotation center.

### 2.2. Acoustic skyrmion bags in moiré superlattices

Building on the twist-engineered acoustic moiré platform and the topological characterization above, we next generate and experimentally observe acoustic skyrmion bags by superposing two skyrmion lattices at an excitation frequency of 2.37 kHz, with $(\varphi_{0,1}, \varphi_{1,1}, \varphi_{2,1}) = (0, 0, 0)$ and $(\varphi_{0,2}, \varphi_{1,2}, \varphi_{2,2}) = (0, 0, 0)$. **Figure 2a** shows the geometrical construction of the twisted bilayer configuration used in our experiments. Two skyrmion lattices are overlaid on the same plane, with a relative rotation angle $\alpha$ between them about a chosen rotation center (red cross). The two twist angles used in Figure 2 correspond to commensurate rotations of a hexagonal lattice, i.e., angles at which the overlaid lattices form well-defined periodic moiré patterns. The smaller angle $\alpha = 13.2°$ produces a multi-skyrmion bag, whereas the larger angle $\alpha = 21.8°$ yields the single-skyrmion limit corresponding to a skyrmionium.

Figure 2b,f show the theoretically calculated out-of-plane velocity fields $v_z$ at the two twist angles. For $\alpha = 13.2°$, the $v_z$ distribution exhibits a large moiré cell containing multiple skyrmions. For $\alpha = 21.8°$, the moiré period is substantially reduced and the enclosed region shrinks to a single skyrmion. The experimentally measured fields in Figure 2c,g are in good agreement with these theoretical predictions. In Figure 2c and Figure 2g, the out-of-plane component is shown as a color map, with the in-plane velocity field indicated by vectors. Figure 2d,h show the corresponding distributions of out-of-plane amplitude (brightness) and in-plane orientation (color). These results show that the bag skyrmion and inner skyrmions share the same in-plane winding, but exhibit different out-of-plane polarity, thereby yielding a skyrmion number of opposite sign for the bag skyrmion. Across both twist conditions, the integration boundaries used to extract $S_{cluster}$ and $S_{bag}$ in Figure 2e,i are defined directly from the measured $v_z$ field: the inner boundary is placed at a local minimum of $v_z$, and the outer boundary at a local maximum. This procedure separates the enclosed skyrmion cluster from the surrounding compensating bag skyrmion and enables a robust determination of the composite topological charges. We obtain $S_{cluster} \approx 7$ and $S_{bag} \approx 6$ for $\alpha = 13.2°$, and $S_{cluster} \approx 1$ and $S_{bag} \approx 0$ for $\alpha = 21.8°$, consistent with a multiskyrmion bag and a skyrmionium, respectively. Together, these results demonstrate that twisting two acoustic skyrmion lattices provides a controllable route to constructing composite topological textures. By selecting appropriate commensurate rotation angles, the number of enclosed skyrmions can be deterministically set, enabling the realization of skyrmion bags containing multiple skyrmions as well as their single-skyrmion limit.

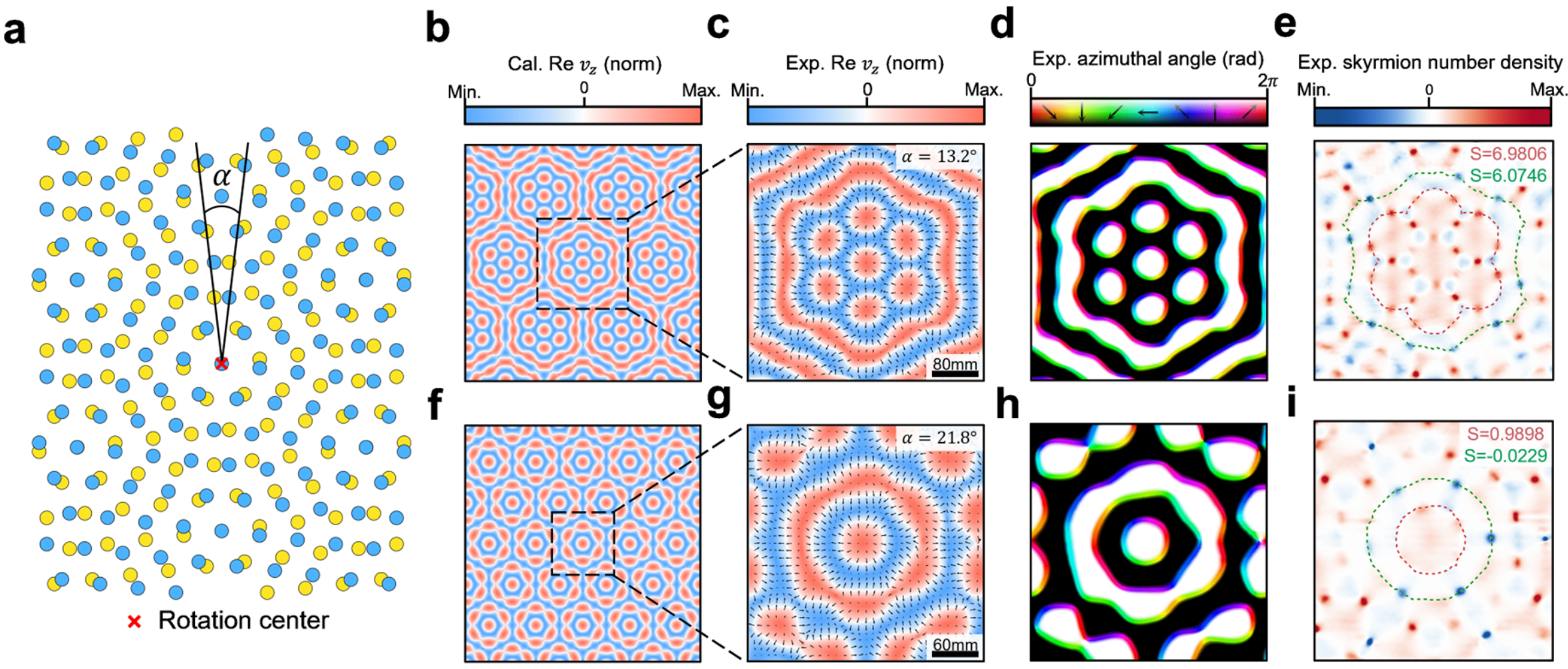

**Figure 2. Experimental observation of acoustic skyrmion bags.** (a) Moiré superlattice formed by overlaying two hexagonal lattices with a relative twist angle $\alpha$. The rotation center is indicated by a red cross at the center of a lattice cell. (b-i) Calculated and experimental results of skyrmion bags with $N = 7$ (top) and $N = 1$ (bottom). (b, f) Calculated out-of-plane component $v_z$ of the acoustic moiré superlattices. (c, g) Measured out-of-plane component $v_z$ (color maps) with the in-plane acoustic velocity fields represented by vectors. The rotation angles $\alpha$ are indicated in the top-right corners, respectively. (d, h) Measured acoustic velocity field distributions, where brightness encodes the out-of-plane amplitude and color encodes the in-plane orientation. (e, i) Distributions of skyrmion number density $s$. The cluster and bag regions are outlined by red and green dashed contours, respectively, with the corresponding skyrmion numbers $S_{cluster}$ and $S_{bag}$ indicated in the top-right corners.

To further explore the geometric degrees of freedom of twisted acoustic lattices, we shift the rotation center from a lattice site to the midpoint between two skyrmion centers, as shown in **Figure 3a**. Experimentally, this is implemented by setting the standing-wave phases to $(\varphi_{0,1}, \varphi_{1,1}, \varphi_{2,1}) = (0, \pi, \pi)$ and $(\varphi_{0,2}, \varphi_{1,2}, \varphi_{2,2}) = (0, \pi, \pi)$, which translates the two skyrmion lattices such that the metasurface center is aligned with the midpoint between neighboring skyrmion centers. This modification changes the relative alignment of the two skyrmion lattices and lowers the symmetry of the resulting moiré patterns from sixfold to twofold. Figure 3b-i show the calculated and experimentally measured skyrmion bags obtained under this shifted-rotation geometry. For a twist angle of $\alpha = 24.4°$, the measured $v_z$ field exhibits a closed envelope containing two skyrmions. The skyrmion number density distribution extracted from the measured velocity field yields $S_{cluster} \approx 2$ and $S_{bag} \approx 1$. Reducing the twist angle to $\alpha = 20.3°$ enlarges the moiré cell and produces a skyrmion bag containing four skyrmions. The skyrmion number density map gives $S_{cluster} \approx 4$ and $S_{bag} \approx 3$. These results show that shifting the rotation center provides an additional geometric degree of freedom for reshaping the bag envelope and selecting the enclosed skyrmion number. Beyond the bilayer cases, we further realize a skyrmion bag formed by superposing three independently oriented skyrmion lattices, as presented in Figure S2 (Supporting Information). This extension indicates that the moiré-based construction can be generalized beyond bilayer interference to higher-order multicomponent topological assemblies.

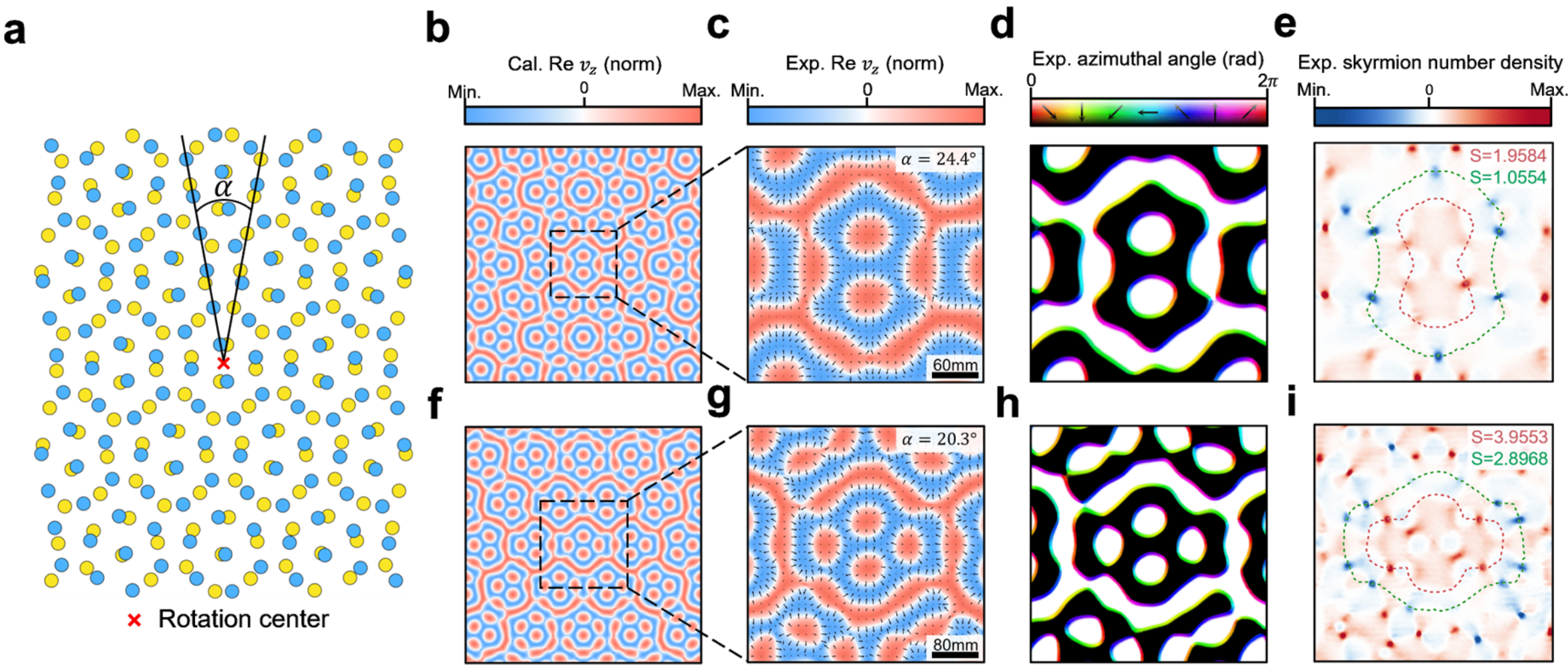

**Figure 3. Skyrmion bags generated by shifting the rotation center.** (a) Moiré superlattice formed by overlaying two hexagonal lattices with a relative twist angle $\alpha$. The rotation center is indicated by a red cross at the midpoint between two neighboring lattice cells. (b-i) Calculated and experimental results of skyrmion bags with $N = 2$ (top) and $N = 4$ (bottom). (b, f) Calculated out-of-plane component $v_z$ of the acoustic moiré superlattices. (c, g) Measured out-of-plane component (color maps) with the in-plane acoustic velocity fields represented by vectors. The rotation angles $\alpha$ are indicated in the top-right corners, respectively. (d, h) Measured acoustic velocity field distributions, where brightness encodes the out-of-plane amplitude and color encodes the in-plane orientation. (e, i) Distributions of skyrmion number density $s$. The cluster and bag regions are outlined by red and green dashed contours,

respectively, with the corresponding skyrmion numbers $S_{cluster}$ and $S_{bag}$ indicated in the top-right corners.

### 2.3. Robustness of acoustic skyrmion bags

To evaluate the defect tolerance of the acoustic skyrmion bags, we introduce controlled defects by inserting rubber plugs into the subwavelength apertures of the metasurface. These defects scatter the SSAWs and perturb the acoustic field within the $30cm \times 30cm$ window where the skyrmion bags are formed (see **Figure 4a**). We vary the defect number $N_d$ and extract $S_{bag}$ and $S_{cluster}$ by integrating the skyrmion number density over the bag and cluster regions. As a reference, we also evaluate $S_{skyrmions}$ for the seven central skyrmions in an untwisted single-layer skyrmion lattice under the same defect conditions (see Figure S3, Supporting Information).

The comparison in Figure 4b shows that the skyrmion bags retain their composite topology over a finite range of defect densities. For low and moderate defect numbers, the extracted $S_{cluster}$ and $S_{bag}$ remain close to their expected values. At higher defect densities, the single-layer reference shows a more noticeable deviation from the ideal seven skyrmion charge, whereas the enclosed skyrmion cluster in the moiré bag remains comparatively closer to its expected value. This comparison suggests enhanced stability of the enclosed skyrmion cluster relative to the single-layer reference at higher defect densities. The error bars in Figure 4b denote sample standard deviations over different defect realizations. The simulations include $20$ random defect configurations for each $N_d$, whereas the experiments include 4 realizations. Thus, the broader simulation error bars mainly reflect more extensive sampling of configuration-to-configuration fluctuations rather than a direct discrepancy between simulation and experiment. As shown in Figure 4c-f, increasing $N_d$ progressively distorts the bag envelope, and at the largest defect density, neighboring inner skyrmions begin to merge, leading to degradation of the composite topology.

We further assess the effect of twist-angle variations by finite-element simulations with the rotation center held fixed. Over a finite interval of twist angles (from $11°$ to $17.5°$), the skyrmion-bag textures retain their composite topology. Representative field maps show that at smaller $\alpha$, the moiré cell expands and the outer boundary tends to open, whereas at a larger $\alpha$ the inner skyrmions approach each other and begin to merge, reducing the topological charge. (see Figure S4, Supporting Information).

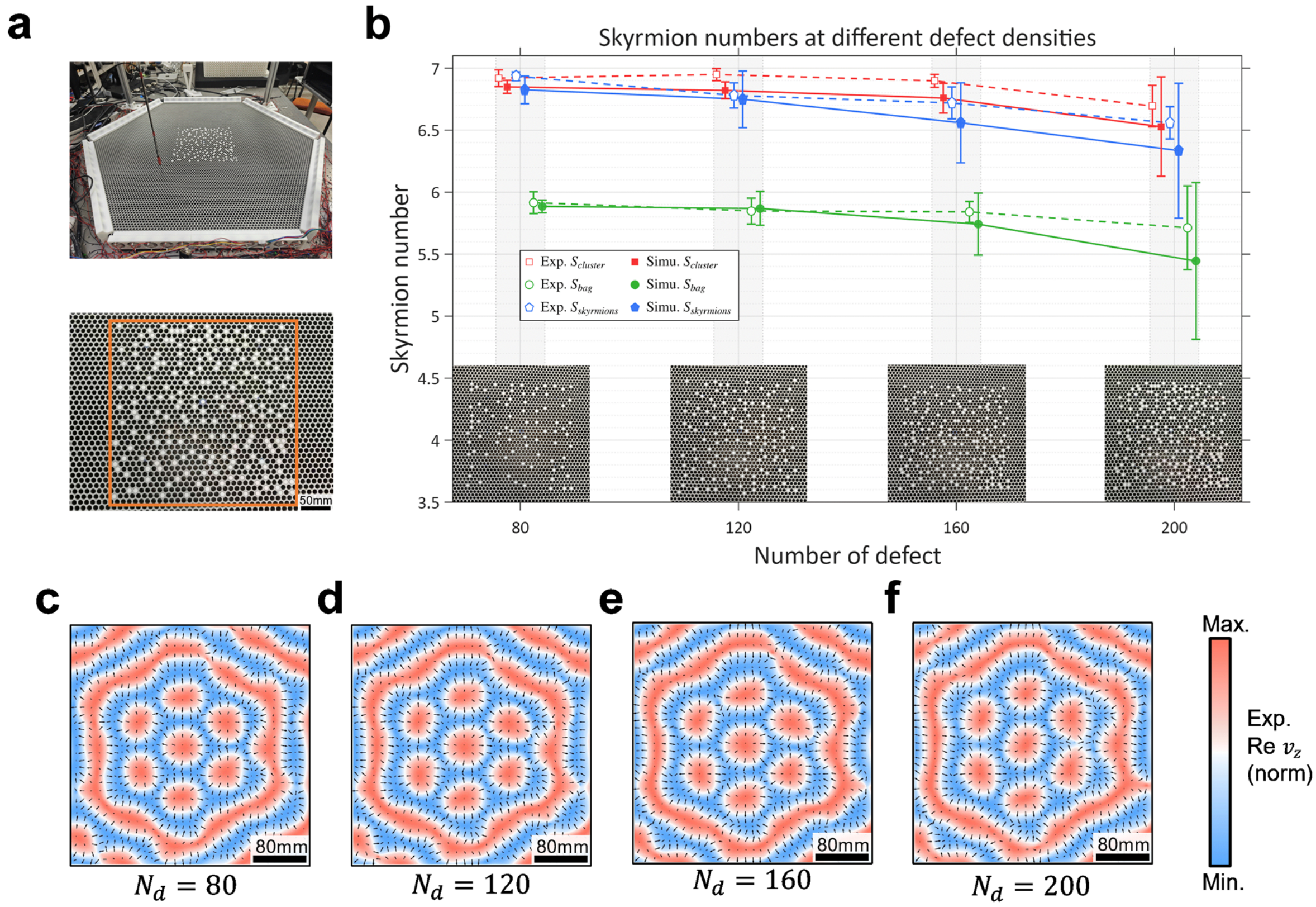

**Figure 4. Robustness of acoustic skyrmion bags.** (a) Local defects are randomly introduced to scatter the SSAWs and distort the acoustic velocity field. (b) Skyrmion numbers $S_{cluster}$ and $S_{bag}$ of the skyrmion bags, together with $S_{skyrmions}$ evaluated from the seven central skyrmions in the single-layer skyrmion lattice, under different defect numbers $N_d$. Markers denote simulation and experimental data, with each marker representing the mean value over 20 simulation realizations or 4 experimental realizations, and error bars indicate the sample standard deviations. Each gray band indicates data points obtained at the same defect number $N_d$ across the six series, and slight horizontal offsets are used to avoid overlap for visual clarity. Insets show the corresponding local defects. (c-f) Measured acoustic velocity field (normalized) distributions of the skyrmion bags under defect numbers $N_d$ (80, 120, 160, and 200). In each figure, the out-of-plane component is visualized using the color plot, and the in-plane acoustic velocity field is represented by vectors.

## 2.4. Phase-driven meron lattices and twist-induced meron clusters

We next use the phase programmability of the SSAW platform to access half-integer meron textures. A phase offset applied to one standing-wave component modifies the relative phasing among the three waves and thereby reshapes the interference pattern. As shown in **Figure 5a**, for a phase shift of $\varphi_0 = -\pi/2$, the measured field develops triangular modulations and exhibits a symmetry reduction from sixfold to threefold. The reconstructed velocity field forms a staggered arrangement of half-integer windings, namely merons and antimerons, and the corresponding skyrmion number density map reveals alternating regions carrying $S = \pm 1/2$ (see Figure 5b-e). These results show that phase control of a single standing-wave component can drive a symmetry-breaking transition from a skyrmion lattice to a meron-antimeron lattice.

Building on the meron-antimeron lattice, we next construct a twisted bilayer meron configuration by superposing two meron lattices on the same metasurface and rotating the lattices by prescribed angles about a chosen rotation center, following the same geometric construction used for the skyrmion bag in Figure 2a. As shown by the experimental results in Figure 5f-i, the meron and antimeron cores become spatially confined within each moiré supercell, forming clustered textures under twist-induced moiré modulation. Because the six outer merons partially merge, we evaluate the net topological charge of the central meron together with the three surrounding antimerons that remain well resolved. The extracted skyrmion number is $S \approx -1$, confirming the formation of a composite meron cluster with a nontrivial net topology. This result shows that the same moiré platform can support both integer and half-integer composite topological textures.

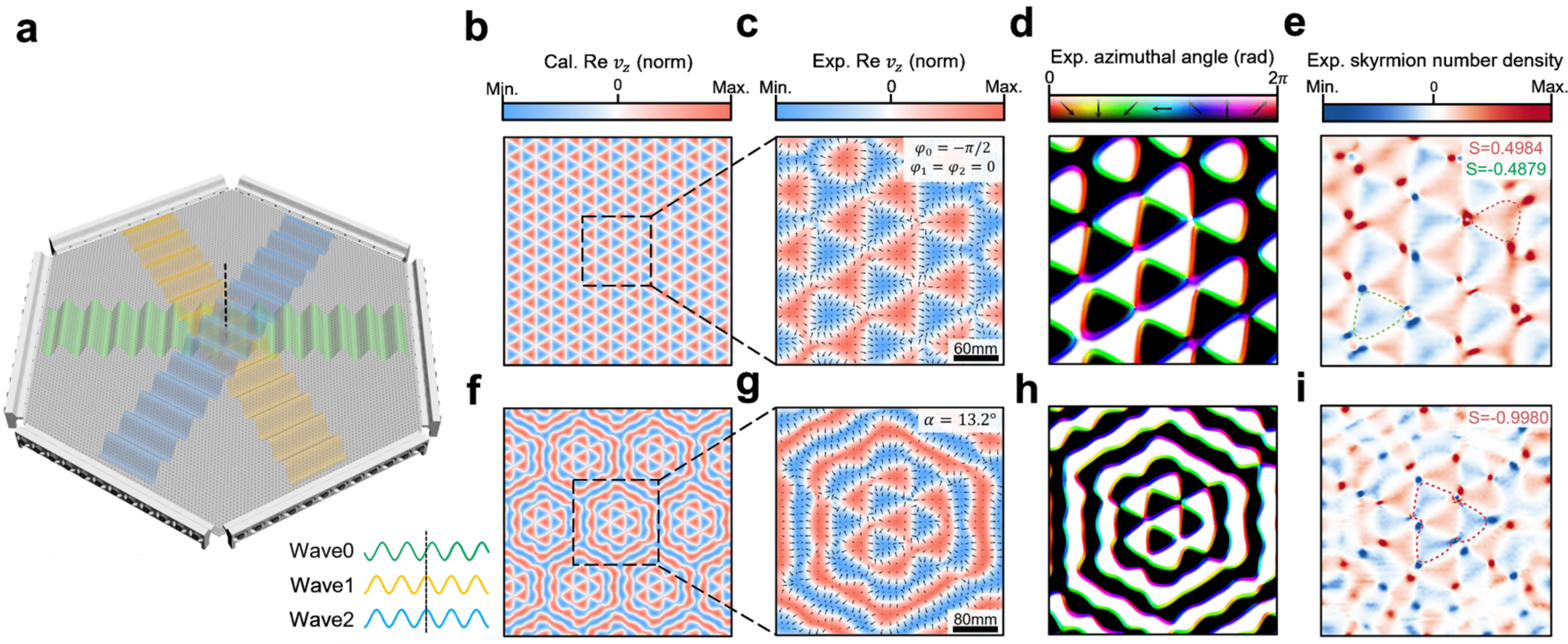

**Figure 5. Phase-driven meron lattices and twist-induced meron clusters.** (a) Three standing waves with spatial phases $\varphi_0 = -\pi/2$ and $\varphi_1 = \varphi_2 = 0$ interfere to generate the meron-antimeron lattice; the vertical dashed line indicates the metasurface center, which serves as the reference for the standing-wave spatial phases. (see the bottom-right inset). (b-i) Calculated and experimental results of the meron-antimeron lattice (top) and the meron cluster (bottom). (b, f) Calculated out-of-plane component $v_z$ of the meron-antimeron lattice and meron clusters superlattice. (c, g) Measured out-of-plane component $v_z$ (color maps) with the in-plane acoustic

velocity fields represented by vectors. The rotation angle $\alpha$ is indicated in the top-right corner of panel (g). (d, h) Measured acoustic velocity field distributions, where brightness encodes the out-of-plane amplitude and color encodes the in-plane orientation. (e, i) Distributions of skyrmion number density $s$. In (e), a meron and an antimeron are outlined by red and green dashed contours, respectively, with the corresponding skyrmion numbers $S_{meron}$ and $S_{antimeron}$ indicated in the top-right corner. In (i), the central meron and three antimerons are outlined by a red contour, with the corresponding skyrmion number indicated in the top-right corner.

**3. Conclusion**

We establish moiré interference as a route to programmable real-space topology engineering in acoustic vector fields. By superposing two skyrmion lattices with a controlled relative rotation on a phase-programmed spoof surface acoustic wave platform, we show that twist-induced moiré interference can assemble composite topological textures in a deterministic and reconfigurable manner. In this framework, the twist angle and rotation center act as independent geometric control parameters that govern the scale, symmetry and internal composition of the resulting moiré textures, enabling acoustic skyrmion bags with controllable topological numbers and geometries. By introducing controlled defects, we find that the skyrmion bags retain their composite topology over a finite range of defect densities. Comparison with an untwisted single layer skyrmion lattice further suggests enhanced stability of the enclosed skyrmion cluster relative to the single-layer reference at higher defect densities. At stronger defect perturbations, distortions of the outer bag boundary and merging of inner skyrmions mark the limits of this stability. Overall, these results show that the moiré construction provides a route to composite acoustic skyrmion textures with tolerance to defect perturbations.

Beyond extending acoustic skyrmionic textures to composite moiré topologies, our results highlight several distinctive features of acoustics for studying real-space topology. Phase-controlled spoof surface acoustic waves provide direct and programmable synthesis of two-dimensional vector fields, while the particle velocity distribution can be reconstructed quantitatively, enabling direct evaluation of topological textures in real space. The coexistence of skyrmion bags, skyrmionium states, meron-antimeron lattices and meron clusters within a single platform further shows that acoustic wave fields can host a broader family of reconfigurable integer and half-integer topological textures than previously accessible. More generally, this work identifies moiré interference as a useful organizing principle for structured acoustic vector fields, and suggests opportunities for topology-guided sound-matter interactions based on reconfigurable near-field localization and closed-loop orientation patterns, including particle trapping, transport and manipulation in engineered acoustic fields.[50-52]

## 4. Methods

### 4.1. Experimental setup

The experimental setup is schematically shown in Figure 1c and Figure S5a (Supporting Information). A total of 72 speakers are arranged as six linear arrays along the six edges of a hexagonal metasurface to excite SSAWs. Each array contains 12 speakers with a spacing of $\Delta x = 4.8$ cm, and each speaker is driven by an independent sound-card channel, allowing individual control of amplitude and phase. By imposing a phase difference $\Delta\phi = k\Delta x sin\theta$ between adjacent speakers, where $k$ is the in-plane wavenumber of the supported SSAW mode, the emitted SSAW component can be steered to an angle $\theta$ relative to the edge normal. By combining plane-wave components with different propagation directions, the speaker arrays synthesize the standing-wave fields required to generate acoustic topological textures with programmable twist angles on the metasurface.

In addition to controlling the propagation directions of the emitted wave components, our setup also enables independent control of their phase offsets. These offsets determine the spatial phases of the synthesized standing waves and thereby allow translation of the generated lattices in real space, which is used to shift the rotation center of the twisted configuration. Moreover, by tuning the spatial phases of the standing-wave components, we can drive controlled transitions between different acoustic topological textures, including skyrmion and meron lattices.

The particle-velocity field above the metasurface is mapped on a scanning plane using acoustic velocity sensors. Because the generated topological textures are three-component particle-velocity vector fields defined over a two-dimensional plane, the three orthogonal components, $v_x$, $v_y$, and $v_z$, are measured in three independent scans over the same region of interest. The component maps are then combined to reconstruct the full vector field, with a synchronized reference signal ensuring phase-consistent acquisition. Figure S5b (Supporting Information) shows the velocity sensor, which converts thermal-field variations induced by the acoustic particle velocity into electrical signals and thereby measures the particle-velocity component perpendicular to the wire direction.

### 4.2. Simulation procedure

All numerical simulations are performed in COMSOL Multiphysics using the Pressure Acoustics, Frequency Domain interface. The perforated metasurface is modeled as an air domain containing an array of cylindrical air regions representing the subwavelength apertures. Each cylinder has one open face, while Sound Hard Boundary conditions are applied to the remaining surfaces. To account for dissipation inside the apertures, the Narrow Region Acoustics condition is applied within the cylindrical hole regions. The air domain above the metasurface is enclosed laterally by a hexagonal boundary, and Plane Wave Radiation conditions are imposed on the six side faces.

Defects corresponding to apertures blocked by rubber plugs are modeled through COMSOL-MATLAB coupling by randomly masking selected cylindrical air regions. Twist-angle variations are simulated by changing the emission angles of the plane-wave components imposed at the radiation boundaries, thereby generating particle-velocity field distributions for different twist angles. For each defect realization and twist-angle setting, the skyrmion numbers $S_{bag}$ and $S_{cluster}$ are calculated from the simulated particle-velocity vector field, allowing us to quantify the robustness of the skyrmion bags against defects and twist-angle variations.

**Acknowledgements**

This work was supported by the National Key R&D Program of China (Grants No. 2023YFA1406904 and 2021YFB3801801), the National Natural Science Foundation of China (Grants No. 52250363 and 52203358), and the Natural Science Foundation of Jiangsu Province (Grants No. BK20232048 and BK20233001). We also acknowledge the support of Singapore Ministry of Education (MOE) AcRF Tier 1 grants (RG157/23 & RT11/23), Singapore Agency for Science, Technology and Research (A*STAR) MTC Individual Research Grants (M24N7c0080), and Nanyang Assistant Professorship Start Up grant.

**Conflicts of Interest**

The authors declare no conflicts of interest.

**Data Availability Statement**

The code used to generate and analyze the data in this study is available from the corresponding authors upon reasonable request.

**Twistsonics: engineering acoustic topological textures in moiré sound lattices**

**Table of Contents**

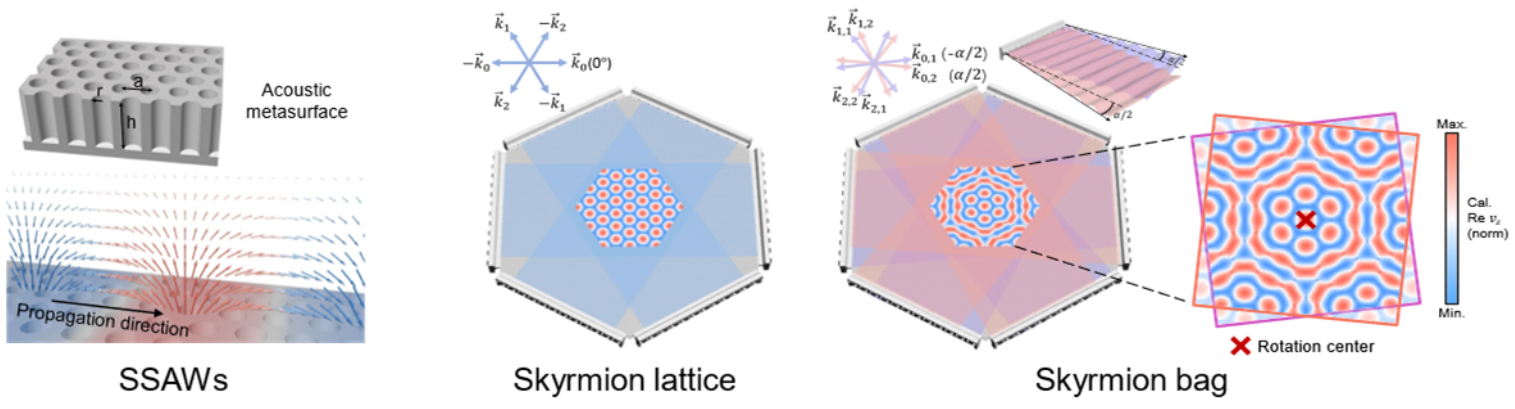

Moiré interference of spoof surface acoustic waves enables real-space engineering of acoustic topological textures. Supported by a perforated metasurface, phase-controlled wave fields assemble skyrmion bags with tunable geometry and topological number, while retaining tolerance to controlled defects and enabling transformations to meron lattices. These results establish a reconfigurable framework for engineering robust real-space topology in acoustic wave fields.